\documentclass[letterpaper,10pt]{article} 

\usepackage{opticameet3} 

\newcommand\authormark[1]{\textsuperscript{#1}}

\usepackage{amsmath,amssymb}
\usepackage[colorlinks=true,bookmarks=false,citecolor=blue,urlcolor=blue]{hyperref} 
\usepackage{color}
\usepackage{xcolor}
\usepackage{graphicx}
\usepackage{tikzscale}
\usepackage{tikz}
\usepackage{caption}
\usepackage{subcaption}
\usepackage{verbatim}
\usepackage{multirow}
\usepackage{wrapfig}
\usepackage{setspace}

\newcommand{\margin}{\ensuremath{\Delta_{\text{SNR}}}}
\newcommand{\SNReol}{\ensuremath{\text{SNR}_{\text{full}}}}
\newcommand{\SNRprob}{\ensuremath{\text{SNR}_{\text{current}}}}

\newcommand{\Pch}{\ensuremath{\text{P}_{\text{ch}}}}

\begin{document}

\title{Experimental Demonstration of ML-Based DWDM System Margin Estimation}


\author{Jasper~Müller\authormark{1,2,*}, Frank~Slyne\authormark{3}, Kaida~Kaeval\authormark{1}, Sebastian~Troia\authormark{4}, Tobias~Fehenberger\authormark{1}, Jörg-Peter~Elbers\authormark{1},
Daniel C. Kilper\authormark{3},
Marco Ruffini\authormark{3}, and Carmen~Mas-Machuca\authormark{2}}

\address{\authormark{1}ADVA, Fraunhoferstr. 9a, 82152 Martinsried/Munich, Germany\\ 
\authormark{2}Chair of Communication Networks, Technical University of Munich,  Munich, Germany\\
\authormark{3}Connect Research Centre, School of Computer Science and Statistics, Trinity College Dublin\\
\authormark{4}Dipartimento di Elettronica, Informazione e Bioingegneria, Politecnico di Milano, Milan 20133, Italy}
\email{\authormark{*}jmueller@adva.com} 
\vspace{-1.2\baselineskip}
\begin{abstract}
SNR margins between partially and fully loaded DWDM systems are estimated without detailed knowledge of the network. The ML model, trained on simulation data, achieves accurate predictions on experimental data with an RMSE of 0.16~dB.\\
\end{abstract}
\vspace{-0.1\baselineskip}
\section{Introduction}
\vspace{-0.5\baselineskip}
With the rapid growth of traffic demands in optical networks, the optimized use of existing network infrastructure is becoming increasingly important. When deploying a coherent optical channel over a lightpath, the choice of configuration in terms of symbol rate and modulation format is typically based on the QoT estimation provided by approximate analytical models such as the Gaussian noise (GN) model \cite{GN}. Traditionally, large margins are subtracted from the QoT estimate when choosing a configuration in order to ensure stable operation, accounting for the ageing of equipment, increased non-linear interference (NLI) due to a higher channel load and inaccuracy of the QoT estimation due to lack of parameter knowledge and the approximate nature of the GN model.

As such, probing a network using performance monitoring data of a deployed transponder has proved to be an effective solution for QoT estimation in optical networks \cite{KaevalJOCN}, especially in the optical-spectrum-as-a-service (OSaaS) scenario \cite{KaevalECOC} where part of the spectrum of a network is leased from an operator.
The main limitation of channel probing is that it characterizes the channel performance only in the current state of the network. Hence, a margin has to be added to account for increased NLI due to a later higher spectrum load.
This fully loaded system margin has a substantial impact on the overall margin with common values of 1.5 to 3~dB \cite{NLImargin}. Assuming perfect knowledge of the physical-layer network parameters, the fully loaded system margin can be accurately determined using analytical models such as the GN model. In production networks, however, uncertainty in the knowledge of these parameters influence the model's accuracy and thus the computed margins \cite{Pointurier:qot}. Machine learning (ML) has been proposed in order to improve the accuracy of QoT estimates in the presence of parameter uncertainty \cite{Lonardi}.
ML-based QoT estimation is an active area of research and several approaches have been proposed \cite{Pointurier:qot}, mostly validated with simulations and requiring detailed information on physical parameters of the network.

In this work, we present an ML model for fully loaded system margin estimation based on limited knowledge of the network. The model is trained on a simulated data set generated with the GN model. Using only a limited input feature set, our proposed model is able to accurately predict the fully loaded system margins with a root-mean-square error (RMSE) of 0.15~dB. Moreover, we demonstrate that the model adapts well to a real optical network by exploiting experimental data that have been collected in the CONNECT OpenIreland Testbed \cite{OpenIreland}. For this work we have automated the testbed setup used to derive OSaaS system margins in previous work \cite{KaevalECOC}. Results show highly accurate predictions of fully loaded system  margins with an RMSE of 0.16~dB.

\vspace{-0.5\baselineskip}
\section{Machine-Learning Model for Fully Loaded System Margin Estimation}\label{sec:sim_results}
\vspace{-0.3\baselineskip}


We model the fully loaded system margin estimation task as a regression problem, which consists of predicting this margin without detailed knowledge of all physical link parameters. The input to the regression algorithm consists of a reduced set of features that characterize the lightpath itself (e.g. channel launch power, channel central frequency, number of spans) and its spectral characteristics (e.g., the overall spectral occupation of the traversed links and the features of the spectrally adjacent lightpaths). We exploited a Bayesian Ridge Regression \cite{BayesianRidge} model performed on all polynomial combinations of the feature set up to degree four. We aim at addressing the following research questions: 
\textit{Q1)} Are we able to train a fully loaded system margin prediction model based on a reduced set of input features? 
\textit{Q2)} Can we transfer the trained model to a real production network and get accurate predictions?

We have generated a simulation data set of over 100,000 data points using the GN model. The parameter space covers DWDM links with constant optical power spectral density at the input, considering channels of 35 up to 69~GBd of QPSK/16/32/64QAM formats and links of 2 to 30 spans with equal span length between 60 and 120 km. The channel launch power for a 35~GBd channel ranges between $-3$ and 0~dBm. For each data point, we have computed the SNR of the channel under test (CUT) by considering a random partially filled spectrum (\SNRprob) and the entirely filled C-band (\SNReol). The fully loaded system margin is defined as $\margin=\SNRprob-\SNReol$.

The considered reduced feature set contains only \SNRprob, the channel launch power (\Pch), the center frequency of the CUT, the number of spans of the link as well as the fraction of the spectrum filled at the time of channel probing. This feature set exploits the information gathered with channel probing and represent impactful information on the network without requiring knowledge on physical parameters.
For data preprocessing, the input features are scaled to zero mean and unit variance. The data set was divided into training (70\%), validation (10\%) and test (20\%) data sets. A model consisting of a Bayesian Ridge Regression \cite{BayesianRidge} performed on all polynomial combinations of the feature set up to degree four was found sufficient and selected for this study. A detailed evaluation and comparison of different ML regression models for the task of fully loaded system margin estimation with the given feature set has been left for future work.

\begin{figure}[ht!]
\centering
\subfloat[Estimation error]{
	   \centering
	   \includegraphics[width=0.24\columnwidth]{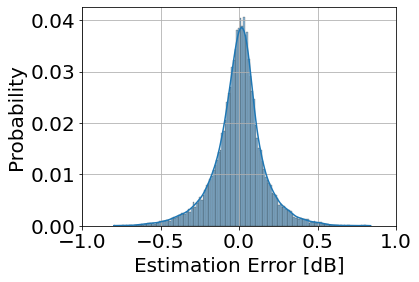}\label{fig:simulation_resultsa}
	   }%
\subfloat[Margin over frequency]{
	   \centering
	   \includegraphics[width=0.24\columnwidth]{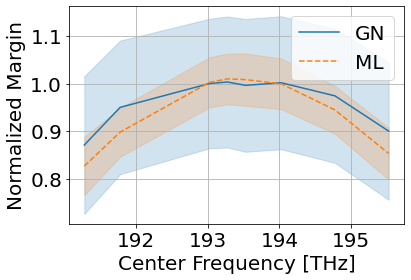}\label{fig:simulation_resultsb}}%
\subfloat[Margin over filled spectrum]{
      \centering
	   \includegraphics[width=0.24\columnwidth]{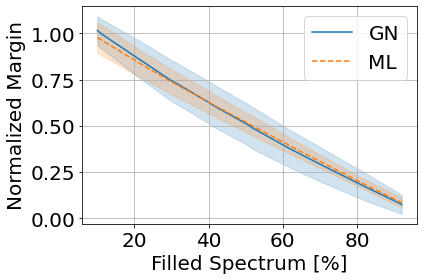}\label{fig:simulation_resultsc}
	   }%
\subfloat[RMSE over granularity]{
	   \centering
	   \includegraphics[width=0.24\columnwidth]{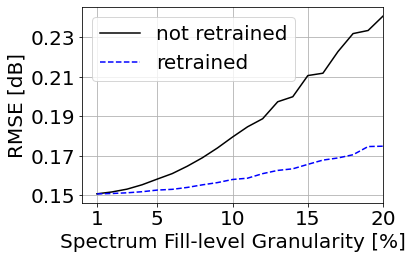}\label{fig:spectrum_granularity}
	   }%
\vspace{-1.3\baselineskip}
\caption{Model evaluation on simulation data. (a) Estimation error distribution, (b) Center frequency impact on margin including standard deviation over different spectrum realizations and topologies, (c) Filled spectrum impact on margin including standard deviation and (d) RMSE depending on the granularity of the filled spectrum feature.}
\end{figure}

In the following, we compare our proposed ML model to the GN model. The distribution of the margin estimation error is shown in Fig.~\ref{fig:simulation_resultsa}. The model achieves an RMSE of 0.15~dB. Fig.~\ref{fig:simulation_resultsb} shows the impact of the center frequency on the margin computed by GN (blue) and predicted by the ML model (orange, dashed). For comparability of predictions on different link topologies, the margin was normalized based on the GN computation. The mean (line) as well as the standard deviation (area) of the normalized margin distribution at a given center frequency are shown for GN computation as well as ML estimation, sampled over different topologies and spectrum realizations of a spectrum filled by 30\%. For each center frequency, a range of normalized margin values is observed since the fully loaded system margins depend on the respective spectral occupation. For example, a spectrum realization where all interfering channels (INTs) are far from the CUT will lead to a higher probing SNR and thus more required margin than a realization where the INTs are close to the CUT. In analogy, longer link lengths will lead to increased NLI and higher required margins. We observe in Fig.~\ref{fig:simulation_resultsb} that the ML model learns the expected relation of an increasing margin needed for channels closer to the center of the C-band and shows a good fit with the GN predictions. In Fig.~\ref{fig:simulation_resultsc}, the impact of the filled bandwidth feature is shown in the same way, sampled over different topologies, center frequencies and spectrum realization for the given percentage of spectrum filled. 
The negative linear correlation that is observed between predicted margin and the filled bandwidth feature allows to further reduce the required knowledge of the network, as the precise value for relative portion of the C-band filled at the time of channel probing can be replaced by a rough upper bound estimation for a more conservative margin prediction. Fig.~\ref{fig:spectrum_granularity} shows the relation between the granularity of considered values and the RMSE of the estimation for the model trained on exact values and a model retrained for the given granularity. It can be seen that while the RMSE increases by around 0.1 dB for the original model, a retrained model stays below 0.18~dB RMSE when considering the filled spectrum feature in a granularity of 20\% steps.


\vspace{-0.8\baselineskip}
\vspace{-2.5\baselineskip}
\section{Experimental Data Collection}
\vspace{-0.5\baselineskip}



\begin{figure}[ht!]
   \includegraphics[width=\columnwidth]{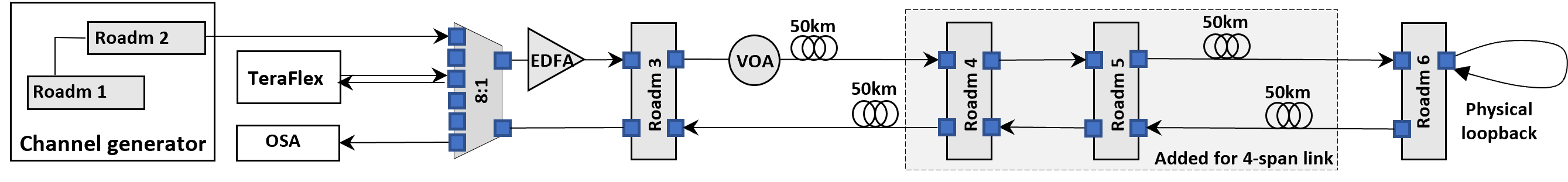}
\vspace{-2.1\baselineskip}
\caption{Experimental link setups.}
\label{fig:experimental_setup}
\end{figure}

The experiment was setup in the OpenIreland Testbed \cite{OpenIreland}. Fig.~\ref{fig:experimental_setup} shows the implemented network topology. The two considered experimental links consist of two and four 50~km spans, respectively, traversing ROADMs in between. A physical loopback at the far-end ROADM is used to extend the transmission distance and simplify data collection. The setup contains the commercially available ADVA TeraFlex \cite{TeraFlex} as CUT, configured to a 200 GBit/s 16-QAM channel in a 50 GHz slot. The interfering channels are generated using shaped amplified spontaneous emissions (ASE) noise in ROADM 1, resulting in a continuous 50-GHz grid of 95 ASE-loaded channels. This grid is processed in ROADM 2 where an arbitrary combination of channels can be let through to the line. The CUT and the noise channels are combined in the 8-port splitter/combiner (8:1), entering a ROADM 3 for power equalization and amplifications. All amplifiers in the schema are set in constant-gain mode. A variable optical attenuator (VOA) in front of the first span is used to tune \Pch~between -3 and 0~dBm.
A set of automated functions was used to vary the location of the CUT within the 95 channel grid, turn the ASE channels on and off, and vary \Pch~per channel.
For the 2-span link, performance monitoring data (SNR) was collected for eight center frequencies, eight levels of filled spectrum between 1\% and 100\% and six \Pch~values between 0 and $-3$~dBm. For each parameter combination, the measurement was repeated ten times for different realizations of the random spectrum.
A reduced number of parameters was considered for data generation on the 4-span link.

\vspace{-0.7\baselineskip}
\vspace{-2.3\baselineskip}
\section{ML Model Evaluation on Experimental Data}
\begin{figure}[ht!]
\centering
\subfloat[Estimation error]{
	   \includegraphics[width=0.24\columnwidth]{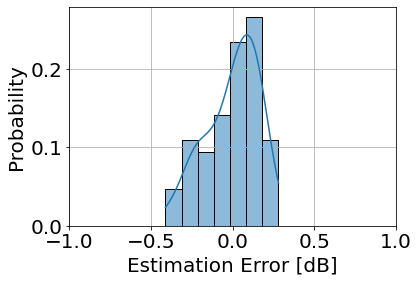}\label{fig:experimental_resultsa}}%
\subfloat[Margin over frequency]{
	   \includegraphics[width=0.24\columnwidth]{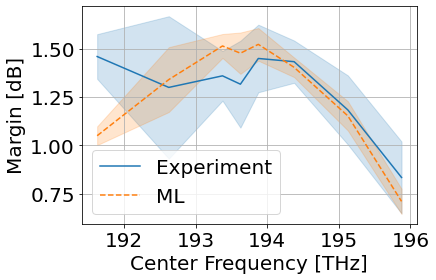}\label{fig:experimental_resultsb}}%
\subfloat[Margin over filled spectrum]{
	   \includegraphics[width=0.24\columnwidth]{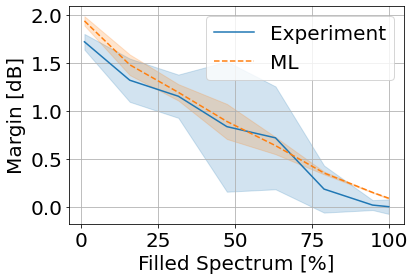}\label{fig:experimental_resultsc}}%
\subfloat[Margin over power]{
	   \includegraphics[width=0.24\columnwidth]{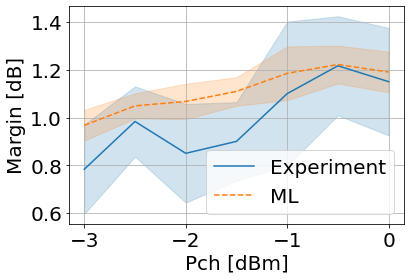}\label{fig:experimental_resultsd}}%
\vspace{-1.3\baselineskip}
\caption{Model evaluation on experimental data showing distributions over different spectrum realizations. (a) Estimation error distribution  and (b), (c), (d) feature impact on margin including standard deviation.}
\end{figure}

\vspace{-0.5\baselineskip}
As explained above, the ML-based fully loaded system margin estimation model was trained on simulation data with a parameter range of typical long-haul networks using a small number of features as selected in Sec.~\ref{sec:sim_results} to reduce the required knowledge on the network topology. Due to constraints in the experimental setup and in order to demonstrate the adaptability of the model, the link length of the setup is shorter than the average link length used during training. Additionally, insertion losses, filtering penalties, frequency-dependent effects and other effects are difficult to model in simulations. We adapt the model to the experimental link by fitting a linear regression on the output of the model. Only five data points from the experimental dataset are shown to provide a stable fit to the link. Without adjustment, the model underestimates the required fully loaded system margins on this experimental link, achieving an RMSE of 0.68 dB. While a retraining of the model on a small experimental data set will be prone to overfitting, the adjustment through linear regression proves effective in re-centering the estimation error distribution (Fig.~\ref{fig:experimental_resultsa}) and reduces the RMSE to 0.16 dB. Fig.~\ref{fig:experimental_resultsb} shows measured and predicted margin over the center frequency for 0~dBm launch power per channel and different spectra with 15\% of the C-band filled. A good fit is observed over most of the spectrum while the measured margin is higher than the predicted margin at the lower end of the spectrum. This could be due to frequency-dependent effects not considered in simulations. In Fig.~\ref{fig:experimental_resultsc} the measured and predicted margins are analysed in relation to the percentage of spectrum filled for 0 dBm launch power per channel and CUT center frequency 193.625~THz. We observe a close fit of the ML predictions to the measurements. The standard deviation of the measured margin is largest around 50\% spectrum filled as the different realizations of the random spectrum have the largest difference in effect on the margin in this parameter range. Fig.~\ref{fig:experimental_resultsd} shows the margin over the launch power per channel for this CUT at 30\% filled spectrum. The ML prediction follows the general trend of lower margins required for lower launch power per channel but does not predict the local minimums at -2 dBm observed in the measurements. The adapted ML model accurately models the general effects of different input features on the fully loaded system margin. The model achieves a comparable accuracy of 0.16 dB RMSE on the reduced dataset for the 4-span link (not shown due to space constraints).
\vspace{-1\baselineskip}
\vspace{-2.4\baselineskip}
\section{Conclusions}
\vspace{-0.2\baselineskip}
We developed an ML model for fully loaded system margin estimation requiring limited information on the network. The model was developed and evaluated on simulation data, showing accurate predictions of the DWDM system margin. An automated experimental setup enabling measurements with varying spectra was presented. The ML model is shown to be easily adaptable to a real link with as little as five data points being required for the adaptation. The model can extend the method of probing the network using performance monitoring data towards fully loaded system margin estimation giving an accurate estimate while not requiring detailed information on the topology and therefore proving viable also for the optical-spectrum-as-a-service use-case.
\\[5pt]
{\scriptsize
\noindent This work has been partially funded by the German Federal Ministry of Education and Research in the CELTIC-NEXT project AI-NET-PROTECT (\textbf{\#16KIS1279K}) and the project 6G-life (\textbf{\#16KISK002}).
Work was also funded by Science Foundation Ireland projects OpenIreland (\textbf{18/RI/5721}) and \textbf{13/RC/2077\_p2}.  \par}
\vspace{-0.8\baselineskip}

\end{document}